# Scalable Production of Highly-Sensitive Nanosensors Based on Graphene Functionalized with a Designed G Protein-Coupled Receptor

*Mitchell B. Lerner[1,2], Felipe Matsunaga[3], Gang Hee Han[1], Sung Ju Hong[1,5], Jin Xi[3], Alexander Crook[1], Jose Manuel Perez-Aguilar[4#], Yung Woo Park[5], Jeffery G. Saven[4], Renyu Liu[3]\*, A.T. Charlie Johnson[1]\**

[1]Department of Physics and Astronomy, University of Pennsylvania, Philadelphia PA 19104, USA

[2]Functional Nano Devices Lab, SPAWAR Systems Center Pacific, San Diego CA 92152 USA

[3]Department of Anesthesiology and Critical Care, University of Pennsylvania, Philadelphia PA 19104, USA

[4]Department of Chemistry, University of Pennsylvania, Philadelphia PA 19104, USA

[5]Department of Physics and Astronomy, Seoul National University, 1 Gwanak-ro, Gwanak-gu, Seoul, 151-747, Korea

ABSTRACT: We have developed a novel, all-electronic biosensor for opioids that consists of an engineered mu opioid receptor protein, with high binding affinity for opioids, chemically bonded to a graphene field-effect transistor to read out ligand binding. A variant of the receptor protein that provided chemical recognition was computationally redesigned to enhance its solubility and stability in an aqueous environment. A shadow mask process was developed to fabricate arrays of hundreds of graphene transistors with average mobility of ~1500 $cm^2$ $V^{-1}$ $s^{-1}$ and yield exceeding 98%. The biosensor exhibits high sensitivity and selectivity for the target naltrexone, an opioid receptor antagonist, with a detection limit of 10 pg/mL.

Graphene field effect transistors (GFETs) hold tremendous promise for use as biosensor transduction elements due to their high carrier mobility and low noise.[1,2] There is a need for scalable, reproducible fabrication of GFET arrays that maintain the high graphene mobility, which can be significantly degraded[3-6] due to contamination by conventional lithographic processing.[7,8] It is also desirable to develop approaches where a biological chemical recognition element, e.g., a transmembrane G-protein coupled receptor, is redesigned for optimized integration with nanotechnology while retaining the protein's native structure, sensitivity and selectivity. We developed a class of opioid biosensors that integrates high quality GFETs for transduction of ligand binding with a computationally designed water-soluble variant of the human mu-opioid receptor that marries functionality of the parent membrane protein with superior properties for production and handling. The fabrication method is scalable with high yield, and sensors exhibit high sensitivity and specificity for the target. The approach is applicable to a variety of engineered proteins. The work represents progress toward enhanced methodologies for detection of small molecules at pg/mL concentrations, precise monitoring of administration of such molecules, and high-throughput testing of the affinities of pharmaceutical compounds for proteins of interest with direct electronic read out.

Large-area graphene grown by chemical vapor deposition (CVD) is appropriate for scalable integration of graphene devices. Care must be taken to avoid lithographic contaminants, which strongly influence sensor responses.[9] To this end, GFET fabrication was based on an approach where graphene is patterned during the transfer process rather than by post-transfer photolithography. Figure 1a is a schematic of the process. Gold lines separated by ~ 150 μm-wide parallel strips were deposited through a mechanical mask onto a graphene layer on its copper growth substrate (Fig. 1b). When a "bubbling" method transfer[10] was performed, graphene covered by gold remained pinned on the copper foil, while uncovered regions were transferred onto an oxidized Si wafer with pre-fabricated source and drain contacts (Fig. 1c). Though limited to graphene "ribbon" widths greater than ~100 μm, the process enabled parallel fabrication of hundreds of GFETs, with excellent reproducibility. To quantify GFET quality, hundreds of devices were fabricated and characterized by electrical transport and Raman spectroscopy. The average mobility, $\mu$, and Dirac voltage, $V_D$, for 212 devices were $\mu = 1496$ cm$^2$ V$^{-1}$ s$^{-1}$ ± 567 cm$^2$ V$^{-1}$ s$^{-1}$ and $V_D = 15.0$ V ± 5.3 V (Fig. 2b-c). Raman spectroscopy of GFET channels (Fig. S1 of the Supporting Information) revealed D/G intensity ratio less than 0.03, indicative of a low defect density, and 2D/G intensity ratio of ~1.5, typical of monolayer graphene.[11]

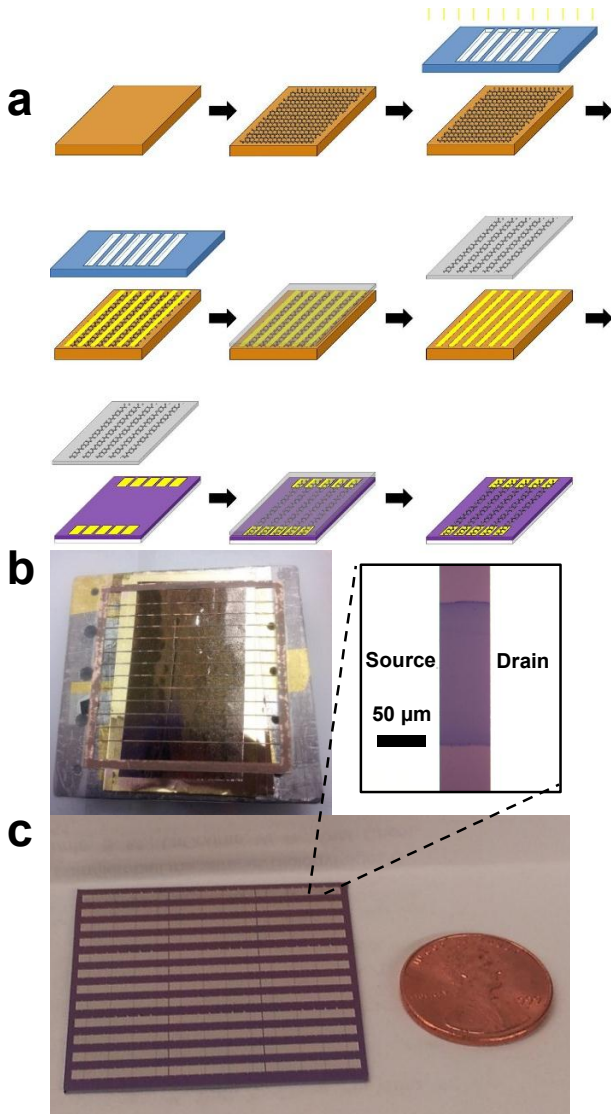

**Figure 1.** Fabrication process for high quality graphene field effect transistors (GFETs). a, Schematic of the fabrication process (see Methods and main text for description). b, Copper foil shadow mask placed in contact with graphene on catalytic copper foil. Narrow regions of graphene protected by the mask are eventually transferred onto source and drain contacts to form the transistor channel. c, Example of a GFET device made by transferring graphene stripes onto pre-fabricated electrodes, and photograph of an array of 192 GFET devices, with ~ 99.5% device yield.

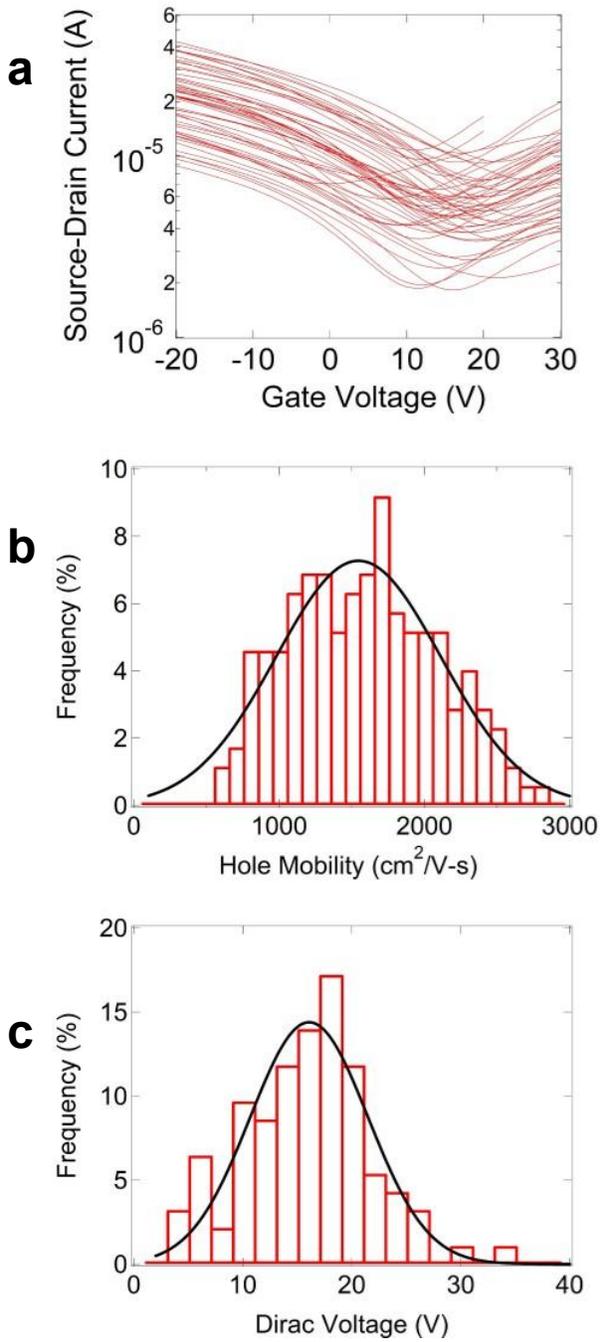

**Figure 2.** Performance characteristics of graphene field effect transistors (GFETs). a, Representative set of 50 I-$V_g$ curves, demonstrating the uniformity of the electrical characteristics. b, c Histograms of GFET mobility and Dirac voltage, along with Gaussian fits (black curves).

G protein-coupled receptors (GPCRs) are a family of transmembrane proteins involved in activating intracellular signal transduction pathways.[12] GPCRs bind a variety of target ligands, from small molecules to large proteins. They are involved in the transmission pathways of numerous diseases and are the target of more than 40% of modern pharmaceuticals.[13] The μ opioid receptor (MUR) is a GPCR involved in pain and reward signaling pathways with high binding affinities for opioids, e.g., β-endorphin, heroin, morphine, hydrocodone, and fentanyl.[14] Generally, GPCRs are unstable when removed from their hydrophobic membrane environment, resulting in denaturation, loss of functionality, and aggregation. Moreover, high-yield heterologous expression and efficient purification of GPCRs remains challenging. Exterior amino acid residues of GPCRs in the transmembrane region are largely hydrophobic, which can impede expression and isolation. By computationally redesigning these residues to be hydrophilic, GPCRs and other membrane proteins can be identified that are expressed in large quantity in *E. coli*, are structurally stable outside of a membrane, and exhibit functionally related properties.[15-19] A designed, water-soluble variant of human MUR can be expressed in *E. coli* and retains opioid affinities comparable to the wild type receptor.[19] No membranes or membrane surrogates are required. The combination of these advances in obtaining functional forms of receptor proteins (GPCRs) that can be manipulated outside biomembranes and the GFET fabrication procedure outlined above opens a route to highly sensitive nanosensors, where the recognition element is essentially the biological receptor protein.

In this work, we demonstrated a bioelectronic GFET nanosensor based on a solubilized MUR variant, and we used it to detect naltrexone, an opioid receptor antagonist, at concentrations as low as 10 pg/mL with excellent specificity. The graphene functionalization scheme presented here can be readily applied to other proteins; the work reveals a new family of biosensors that combine the functional properties of GPCRs with the environmental sensitivity of graphene for tailored and targeted chemical detection.

GFET arrays of were functionalized with water-soluble MUR using a methodology based on our earlier experiments with exfoliated graphene.[20] To our knowledge, this is the first application of this approach to devices based on large-area graphene. The process began with incubation in a solution of 4-carboxybenzene diazonium tetrafluoroborate, which produces carboxylic acid sites on the graphene that were then activated and stabilized with 1-ethyl-3-[3-dimethylaminopropyl] carbodiimide hydrochloride /sulfo-N hydroxysuccinimide (EDC/s-NHS) in MES buffer. Incubation in a buffer with the water-soluble MUR led to covalent attachment of the designed MUR and the graphene (see Methods for further details). To measure the sensor response, a solution containing a known concentration of naltrexone in buffer was delivered to the sensor and allowed to react for 40 min before being rinsed with DI water and blown dry.

Devices were characterized through the functionalization process by Raman spectroscopy of the GFET channel and Atomic Force Microscopy (AFM). Raman spectra of GFETs after incubation in diazonium salt solution (Fig. 3a) displayed strong increases in the D ("disorder") peak ca. 1360 $cm^{-1}$, consistent with formation of $sp^3$ hybridized sites.[21] AFM showed enhanced binding of water-soluble MUR to the graphene sheet compared to the $SiO_2$ substrate and verified the effectiveness of the attachment chemistry, e.g. 128 proteins bound to 27 $\mu m^2$ of graphene (4.7/$\mu m^2$) and 5 protein-sized features in an area of 9 $\mu m^2$ of substrate (0.55/$\mu m^2$) in Fig. 3b. AFM line scans were used to create a height histogram for immobilized proteins (Fig. 3c), which showed a primary maximum at ~ 4 nm, consistent with the 46 kDa mass and structure of MUR;[22] secondary maxima at 8 and 12 nm were attributed to protein aggregates. To check that proteins were bound to the graphene covalently rather than by non-specific adsorption, the functionalization procedure was performed with the diazonium salt step omitted. In this experiment, the density of non-specifically adsorbed protein on both the graphene and the oxidized silicon substrate was similar to that observed on the bare substrate in Fig 3b (Fig S2 of the Supporting Information).

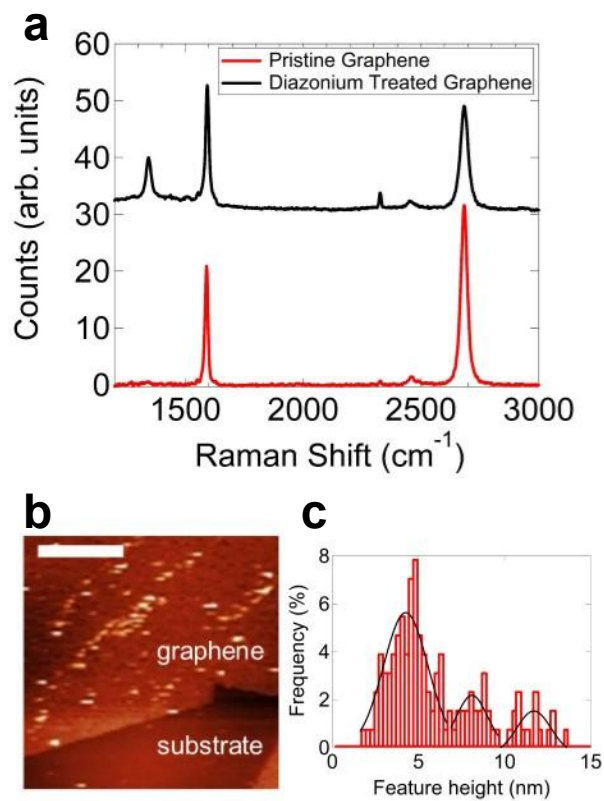

**Figure 3.** Results of characterization by Raman spectroscopy and Atomic Force Microscopy (AFM). a, Raman spectrum of graphene before (red data) and after (black data) exposure to diazonium salt solution. The strongly enhanced D-band (near 1360 cm$^{-1}$) after diazonium treatment indicates the formation of carboxy-benzene sites on the graphene surface. b, AFM image of soluble mu receptor proteins (white dots) decorating the graphene surface. The density of protein molecules is approximately 10 times greater on the graphene as compared to the SiO$_2$ substrate. Scale bar is 2 μm. c, Histogram of the heights of proteins indicating that the 46 kDa mu receptor monomer is ~4 nm tall on the surface, with dimers and trimers of 8 nm and 12 nm respectively.

Samples were characterized after each step of functionalization chemistry and exposure to naltrexone target by measuring the source-drain current as a function of back gate voltage (I-$V_G$ characteristic). The Dirac voltage and carrier mobility were very sensitive to chemical treatment (Fig. 4a). After diazonium treatment, $V_D$ increased by 70-90 V, consistent with the presence of negative charge (carboxylate) near the GFET channel.[23] Additionally, mobility was found to be reduced by ~50%, which was ascribed to carrier scattering by $sp^3$-hybridized sites created on the graphene surface. EDC/s-NHS treatment decreased the Dirac voltage by ~20 V, likely due to neutralization of carboxylate by the s-NHS ester. While MUR attachment did not significantly shift the Dirac point, there was an increase in the device mobility of ~10%.

For naltrexone response experiments, each device was exposed to a single concentration to avoid sample contamination across trials; 15-30 devices were tested against each concentration. A reproducible, concentration-dependent increase in $V_D$ was observed (Fig. 4e). Device-to-device scatter in this signal parameter appeared to be random and uncorrelated with other device properties, i.e., the Dirac voltage and carrier mobility. The data are well fit by a model adapted from the Hill-Langmuir equation that describes equilibrium binding of a ligand by a receptor.[24] The model includes a term that reflects naltrexone binding and an offset parameter Z to account for $\Delta V_D$ observed in the absence of naltrexone:

$$\Delta V_D = A \frac{(c/K_a)^n}{1+(c/K_a)^n} + Z$$

Here *A* is the maximum response with all binding sites occupied, *c* the concentration of the applied naltrexone solution, $K_a$ an effective dissociation constant that describes the naltrexone concentration that produces half occupation of mu receptors, $n$ the Hill coefficient, and $Z$ the offset parameter. The best fit to the data yielded values $A = 9.26 \pm 0.24$ V, $K_a = 7.8 \pm 1.6$ ng/mL, $n = 0.41 \pm 0.03$, and $Z = 0.11 \pm 0.03$ V. During the curve fitting process, $A$ was constrained to be in the range of 8.5 – 10 V based on observed responses, and the other parameters were unconstrained.

The best fit value of the offset parameter $Z = 0.11 \pm 0.03$ V agrees with measured responses of devices exposed to pure buffer ($Z = 0.04 \pm 0.38$ V). The best fit value of $K_a = 7.8 \pm 1.6$ ng/mL for the soluble MUR is within the expected range 1.5 – 100 nM, as reported previously for both the designed and wild-type MUR.[19, 25] Under the reasonable, but not yet tested, assumption of a linear relationship between GFET sensor response and analyte binding, the best fit value $n = 0.41 \pm 0.03$ suggests negative cooperativity in the binding of naltrexone to the GFET biosensor that may be due to protein-protein interactions upon binding or increased charge carrier scattering with increased ligand binding. Similar behavior (n < 1) was observed in biosensors based on protein-functionalized carbon nanotube transistors.[26] These results demonstrate that a collection of 15-30 GFET devices functionalized with the water soluble MUR can distinguish between pure buffer ($\Delta V_D = 0.04 \pm 0.38$ V) and a solution containing naltrexone at a concentration of 10 pg/mL (3 pM; $\Delta V_D = 0.93 \pm 0.34$ V).

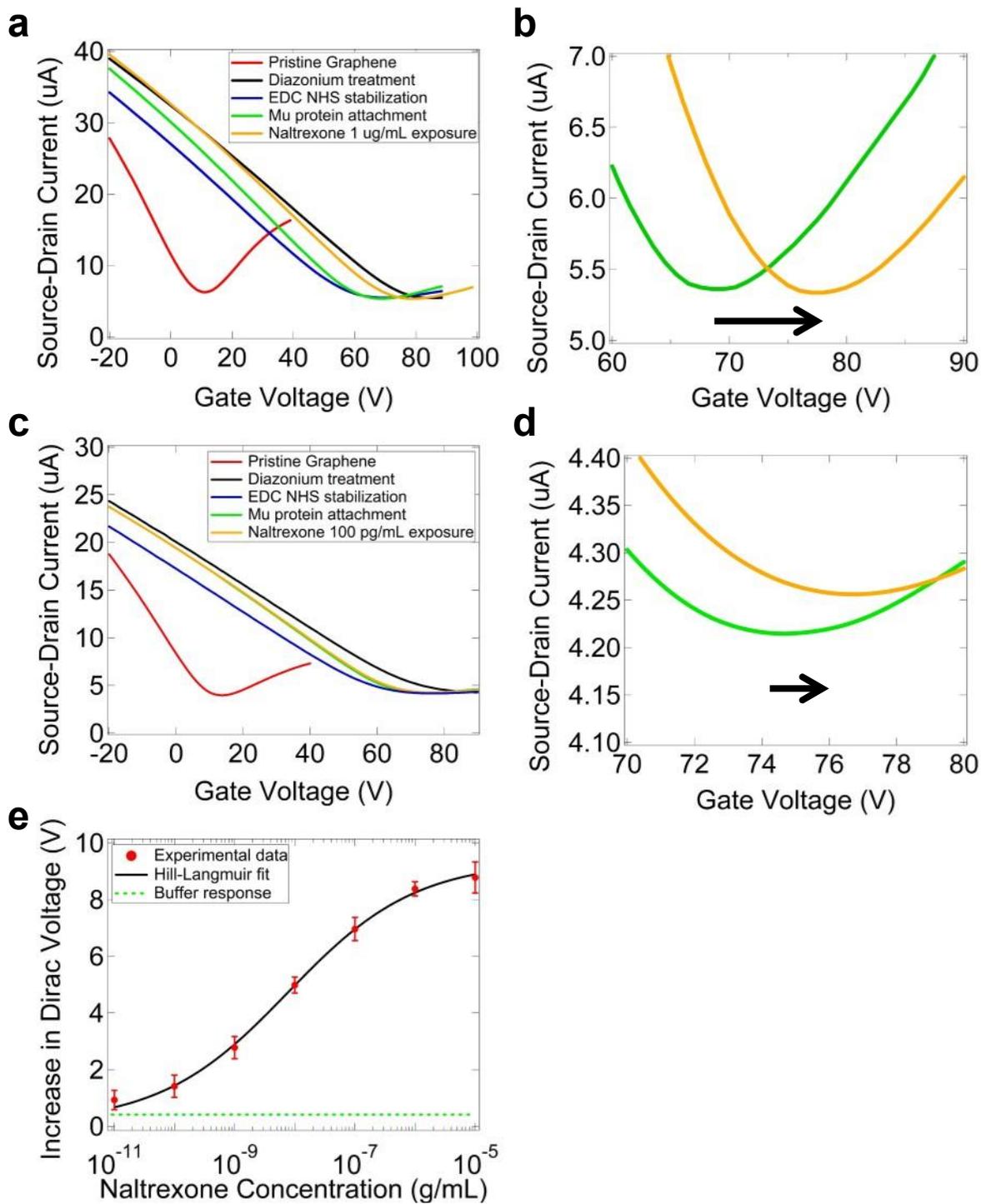

**Figure 4.** Current-gate voltage (I-V$_G$) characteristic measurements after chemical treatment and naltrexone exposure. a, I-V$_G$ plots after successive functionalization steps. After functionalization with the solubilized mu receptor, exposure to a solution of 1 µg/mL naltrexone in buffer leads to an increase in the Dirac voltage of 8.5 V (green curve

to orange curve). b Magnified view of the Dirac voltage increase. c I-$V_G$ plots after successive functionalization steps with the device now exposed to a solution of 100 pg/mL naltrexone in buffer (green curve to orange curve). The Dirac voltage increase is 1.8 V. d Magnified view of this shift in the Dirac voltage. e Sensor response (increase in Dirac voltage) as a function of Naltrexone concentration. The signal is still discernable from the bare buffer response at 10 pg/mL naltrexone. The data are fit to a modified Hill-Langmuir equation (black curve; see main text for details). The buffer response (green line) is defined as the average response plus one standard error of 15 GFETs exposed to pure buffer without protein.

Control experiments were conducted on 12-20 GFET devices per condition to verify that sensor responses reflected specific binding of naltrexone to the soluble MUR. Results are summarized in Table 1. Devices functionalized with water-soluble MUR were exposed to pure buffer with no naltrexone; the sensor response was consistent with zero ($\Delta V_D = 0.04 \pm 0.38$ V). A benzodiazepine receptor antagonist compound, Flumazenil, at a concentration of 10 µg/mL was delivered to the sensors under conditions identical to the naltrexone experiments. Flumazenil is known not to bind to mu receptors,[27] and indeed we did not observe a statistically significant shift in the Dirac voltage ($\Delta V_D = -0.23 \pm 0.43$ V). Experiments were performed where i) the mu receptor addition step was omitted, and ii) where the mu receptor was replaced with an antibody fragment, anti-HER2 scFv. Sensor responses were measured upon exposure to naltrexone at 10 µg/mL, the highest concentration tested. Again, the response was consistent with zero, implying that the MUR was necessary to bind naltrexone and produce the characteristic $\Delta V_D$. These experiments provide strong evidence that sensor responses derived from specific interactions between the water-soluble MUR and the opioid receptor antagonist naltrexone.

**Table 1.** Summary of control experiments performed to test the conclusion that sensor responses derive from specific binding of naltrexone to the mu receptor protein. Quoted errors are the standard error of the mean.

| Sample | Analyte | Average Dirac Voltage Shift (V) |
|---|---|---|
| MUR-GFET | Buffer with no Naltrexone | 0.04 ± 0.38 |
| MUR-GFET | Flumazenil at 10 μg/mL | −0.23 ± 0.43 |
| MUR omitted | Naltrexone at 10 μg/mL | −0.25 ± 0.35 |
| anti-HER2 scfv-GFET | Naltrexone at 10 μg/mL | −0.31 ± 0.48 |
| MUR-GFET | Naltrexone at 10 μg/mL | 8.78 ± 0.55 |

The proposed mechanism for the concentration-dependent $\Delta V_D$ is a conformational change in the variant MUR upon naltrexone binding, as observed for other GPCRs,[28] which alters the electrostatic environment of the GFET ("chemical gating"). A similar chemical gating response has been demonstrated in carbon nanotube FETs.[23] The Raman spectra of GFETs were also sensitive to the presence and concentration of naltrexone, (Figs. S3-4 and Table S1-2 in the Supporting Information), consistent with the fact that Raman spectra are also sensitive to chemical gating.[29, 30] Although a precise quantitative understanding of the mechanism remains to be developed, we believe that the methods presented here could be generalized to create a new generation of manufacturable graphene-based biosensors with the highly sensitive and specific chemical recognition characteristic of GPCRs.

In conclusion, we demonstrated a novel biosensor based on arrays of graphene FETs functionalized with a designed, water-soluble mu receptor protein. Scalable fabrication methods were used to produce large arrays of high quality graphene transistors as evidenced by electronic and Raman characterization. The devices enabled detection of the opioid receptor antagonist naltrexone at concentrations as low as 10 pg/mL. Measured sensor responses over a range of 6 orders of magnitude in concentration (10 pg/mL to 10 µg/mL) were well fit by a model based on Hill-Langmuir binding equation. Control experiments verified that the sensor response derived from specific binding of the mu receptor to naltrexone, indicating that the water-soluble MUR maintains its biologically active analyte binding configuration while covalently bound to graphene. By functionalizing such GFET arrays with multiple selected proteins, it should be possible to create a single integrated biosensor platform for detection of a large number of analytes. Nanoelectronic interrogation of proteins while they are subjected to new pharmaceutical treatments could perhaps serve as a sensitive readout for drug discovery applications.

**Methods**

**Growth of large-area graphene by chemical vapor deposition (CVD).** Continuous (monolayer) graphene sheets were grown on copper foil by atmospheric pressure CVD according to methods we published previously.[31] Briefly, copper foil (Alfa Aesar Item #46365) was placed in a CVD tube furnace and heated to 1057°C under argon and hydrogen forming gas. Immediately upon reaching 1057°C, methane was introduced to the furnace chamber, and the graphene growth proceeded for 30 min. The furnace was then slid downstream while holding the quartz tube fixed so that the copper foil was outside the furnace heating elements, enabling rapid cooling of the copper foil necessary for monolayer graphene formation. Once the chamber cooled to 100°C, the foil was removed from the furnace.

**Fabrication of GFETs by patterned transfer.** After graphene synthesis, the copper growth substrate was placed into close contact with a stripe-patterned mechanical mask and inserted into a thermal evaporator (Fig. 1a). 7 nm Ti/55 nm Au were deposited on the sample, with the shadow mask creating parallel strips of graphene that were not covered by metal (see Fig. S1 in the Supporting Information). Polymethylmethacrylate (PMMA) was then spun over the surface of the copper foil/graphene/gold sample and baked at 100°C for 2 min. Graphene was then separated from the copper foil using the electrochemical "bubble" transfer method. The sample was slowly lowered into a 0.1 M NaOH solution with a 25 V potential difference applied between the copper foil and the solution. Gas bubbles formed at each electrode, which served to separate the graphene from the copper foil. However, graphene areas covered by gold were pinned to the copper foil, and only the uncovered areas were free to come off with the PMMA support layer. This approach enable the transfer of patterned stripes of graphene without the need for conventional photolithography and without introducing any additional chemicals which may cause unwanted doping or contamination. The PMMA/graphene layer was carefully transferred to 2 deionized water baths and then placed onto a Si wafer (300 nm oxide) so that it aligned with pre-fabricated electrodes for source and drain contacts. The PMMA was rinsed off with acetone and the devices were annealed in Ar/$H_2$ for 1 hour at 200°C to remove remaining PMMA. Device yield at this stage typically exceeded 99%, with excellent uniformity.

**Water soluble human MUR engineering, expression and purification.** A variant of water soluble MUR was computationally designed as described previously, wherein exterior, transmembrane residues were targeted and redesigned.[19] The engineered variant was expressed and purified as described previously without further modification.[19] Briefly, the synthetic cDNA encodings of the water soluble MUR were produced by GenScript Inc. (Piscataway, NJ). The sequences were subcloned between the NdeI and XhoI restriction sites of the expression plasmid pET-28b(+) (EMD/Novagen). This cloning strategy resulted in placement of a His-tag at the amino terminus of the protein. *E. coli* BL21(DE3) cells (EMD/Novagen) were used for expression. The protein was purified using the His-tag and confirmed using mass spectrometry.

**Protein functionalization of GFETs and exposure to naltrexone target.** The first step was incubation at 55°C in a solution of a carboxylated diazonium salt, 4-carboxybenzene diazonium tetrafluoroborate (2.0 mg/ mL in DI water). Diazonium salts were synthesized in house according to a published protocol.[32] Carboxylic acid groups from the diazonium functionalization were activated and stabilized with 1-ethyl-3-[3-dimethylaminopropyl] carbodiimide hydrochloride /sulfo-N hydroxysuccinimide (EDC/s-NHS) at an EDC concentration of 9 mg/15 mL MES buffer and NHS concentration of 20 mg/15 mL MES buffer. NHS molecules were displaced by amine groups on the designed MUR protein (3 µg/mL) in buffer (40µM NaPi/ 260µM NaCl / 0.00004% SDS/ 10µM 2-ME / pH 7.0) to form a covalent amide bond between the soluble MUR and the graphene. Finally, an amount of a known opioid receptor antagonist (naltrexone) was delivered to the sensor and allowed to react in a humid environment for 40 min before being rinsed in DI water and blown dry under nitrogen. The time of 40 min was chosen based on a calculation that 15 min would be required for diffusion of naltrexone molecules to the sensor surface and to ensure a binding equilibrium was established at the lowest concentrations tested.

**Acknowledgements**

This work was supported by the National Science Foundation Accelerating Innovation in Research program AIR ENG-1312202 and the Nano/Bio Interface Center NSF NSEC DMR08-32802. Additional support was provided by FAER (Foundation for Anesthesia Education and Research, PI, R.L.), NIH K08 (K08-GM-093115) (PI, R.L.), GROFF (PI, R.L.), and the Department of Anesthesiology and Critical Care at the University of Pennsylvania (PI, R.L.).

J.G.S. acknowledges infrastructural support from the Penn LRSM MRSEC (NSF DMR-1120901). Y.W.P. and S.J.H. acknowledge support from the Leading Foreign Research Institute Recruitment Program (0409-20100156) of NRF and the FPRD of BK21 through the MEST, Korea. We gratefully acknowledge the use of anti-HER2 scFv antibodies that were generously provided by the laboratory of Dr. Matthew Robinson of the Fox Chase Cancer Center.

**Corresponding Authors**

* Correspondence to: RenYu.Liu@uphs.upenn.edu (R.L.) and cjohnson@physics.upenn.edu (A.T.C.J.)

**Present Addresses**

#Present address: Department of Physiology and Biophysics, Weill Medical College of Cornell University, New York, New York, USA

**Author Contributions**

M.L. fabricated the graphene field effect transistor devices, conducted the protein functionalization, and performed all the electrical measurements described in the paper. F.M. expressed and purified the mu receptor proteins. G.H.H. performed graphene growth and assisted with sensor characterization. S.J.H. performed Raman spectroscopy on graphene devices through each step of the functionalization process. J.X. designed and characterized the mu receptor proteins. A.C. grew high quality graphene used in the experiments. J.M.P.-A. designed the water-soluble variant of the receptor. Y.W.P., J.G.S., R.L., and A.T.C.J. designed the experiment. The manuscript was written through contributions of all authors. All authors have given approval to the final version of the manuscript.

**Supporting Information**

Raman spectrum of graphene field effect (GFET) channel region. Protein functionalization control experiment. Use of Raman spectroscopy for sensor readout. This material is available free of charge via the Internet at http://pubs.acs.org.

TOC Image below

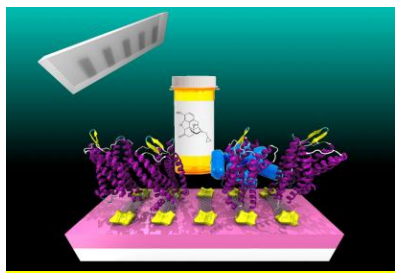

**References.**

1. Chung, C.; Kim, Y.-K.; Shin, D.; Ryoo, S.-R.; Hong, B. H.; Min, D.-H., Biomedical Applications of Graphene and Graphene Oxide. *Acc. Chem. Res.* **2013,** *46*, 2211-2224.
2. Kuila, T.; Bose, S.; Khanra, P.; Mishra, A. K.; Kim, N. H.; Lee, J. H., Recent Advances in Graphene-Based Biosensors. *Biosens. Bioelectron.* **2011,** *26*, 4637-4648.
3. Kumar, S.; Peltekis, N.; Lee, K.; Kim, H.-Y.; Duesberg, G. S., Reliable Processing of Graphene Using Metal Etchmasks. *Nanoscale research letters* **2011,** *6*, 390.
4. Lee, Y.; Bae, S.; Jang, H.; Jang, S.; Zhu, S.-E.; Sim, S. H.; Song, Y. I.; Hong, B. H.; Ahn, J.-H., Wafer-Scale Synthesis and Transfer of Graphene Films. *Nano Lett.* **2010,** *10*, 490-493.
5. Levendorf, M. P.; Ruiz-Vargas, C. S.; Garg, S.; Park, J., Transfer-Free Batch Fabrication of Single Layer Graphene Transistors. *Nano Lett.* **2009,** *9*, 4479-4483.
6. Li, X.; Cai, W.; An, J.; Kim, S.; Nah, J.; Yang, D.; Piner, R.; Velamakanni, A.; Jung, I.; Tutuc, E*., et al.*, Large-Area Synthesis of High-Quality and Uniform Graphene Films on Copper Foils. *Science* **2009,** *324*, 1312-1314.
7. Khamis, S. M.; Jones, R. A.; Johnson, A. T. C., Optimized Photolithographic Fabrication Process for Carbon Nanotube Devices. *AIP Advances* **2011,** *1*, 022106.
8. Ishigami, M.; Chen, J. H.; Cullen, W. G.; Fuhrer, M. S.; Williams, E. D., Atomic Structure of Graphene on Sio2. *Nano Lett.* **2007,** *7*, 1643-1648.
9. Dan, Y.; Lu, Y.; Kybert, N. J.; Luo, Z.; Johnson, A. T. C., Intrinsic Response of Graphene Vapor Sensors. *Nano Lett.* **2009,** *9*, 1472-1475.
10. Gao, L.; Ren, W.; Xu, H.; Jin, L.; Wang, Z.; Ma, T.; Ma, L.-P.; Zhang, Z. T.; Fu, Q.; Peng, L.-M*., et al.*, Repeated Growth and Bubbling Transfer of Graphene with Millimetre-Size Single Crystal Grains Using Platinum. *Nature Communications* **2012,** *3*, 699.

11. Pimenta, M. A.; Dresselhaus, G.; Dresselhaus, M. S.; Cancado, L. G.; Jorio, A.; Saito, R., Studying Disorder in Graphite-Based Systems by Raman Spectroscopy. *Phys. Chem. Chem. Phys.* **2007,** *9*, 1276-1291.
12. King, N.; Hittinger, C. T.; Carroll, S. B., Evolution of Key Cell Signaling and Adhesion Protein Families Predates Animal Origins. *Science (New York, N.Y.)* **2003,** *301*, 361-363.
13. Overington, J. P.; Al-Lazikani, B.; Hopkins, A. L., How Many Drug Targets Are There? *Nat. Rev. Drug Discovery* **2006,** *5*, 993-996.
14. Zhorov, B. S.; Ananthanarayanan, V. S., Homology Models of Mu-Opioid Receptor with Organic and Inorganic Cations at Conserved Aspartates in the Second and Third Transmembrane Domains. *Arch. Biochem. Biophys.* **2000,** *375*, 31-49.
15. Slovic, A. M.; Kono, H.; Lear, J. D.; Saven, J. G.; DeGrado, W. F., Computational Design of Water-Soluble Analogues of the Potassium Channel Kcsa. *Proc. Natl. Acad. Sci. U. S. A.* **2004,** *101*, 1828-1833.
16. Ma, D. J.; Tillman, T. S.; Tang, P.; Meirovitch, E.; Eckenhoff, R.; Carnini, A.; Xu, Y., Nmr Studies of a Channel Protein without Membranes: Structure and Dynamics of Water-Solubilized Kcsa. *Proc. Natl. Acad. Sci. U. S. A.* **2008,** *105*, 16537-16542.
17. Cui, T.; Mowrey, D.; Bondarenko, V.; Tillman, T.; Ma, D.; Landrum, E.; Perez-Aguilar, J. M.; He, J.; Wang, W.; Saven, J. G.*, et al.*, Nmr Structure and Dynamics of a Designed Water-Soluble Transmembrane Domain of Nicotinic Acetylcholine Receptor. *Biochimica et Biophysica Acta - Biomembranes* **2012,** *1818*, 617-626.
18. Perez-Aguilar, J. M.; Saven, J. G., Computational Design of Membrane Proteins. *Structure* **2012,** *20*, 5-14.
19. Perez-Aguilar, J. M.; Xi, J.; Matsunaga, F.; Cui, X.; Selling, B.; Saven, J. G.; Liu, R., A Computationally Designed Water-Soluble Variant of a G-Protein-Coupled Receptor: The Human Mu Opioid Receptor. *PLoS One* **2013,** *8*, e66009.
20. Lu, Y.; Lerner, M. B.; John Qi, Z.; Mitala, J. J.; Hsien Lim, J.; Discher, B. M.; Charlie Johnson, A. T., Graphene-Protein Bioelectronic Devices with Wavelength-Dependent Photoresponse. *Appl. Phys. Lett.* **2012,** *100*, 033110.
21. Eckmann, A.; Felten, A.; Mishchenko, A.; Britnell, L.; Krupke, R.; Novoselov, K. S.; Casiraghi, C., Probing the Nature of Defects in Graphene by Raman Spectroscopy. *Nano Lett.* **2012,** *12*, 3925-3930.
22. Manglik, A.; Kruse, A. C.; Kobilka, T. S.; Thian, F. S.; Mathiesen, J. M.; Sunahara, R. K.; Pardo, L.; Weis, W. I.; Kobilka, B. K.; Granier, S., Crystal Structure of the Mu-Opioid Receptor Bound to a Morphinan Antagonist. *Nature* **2012,** *485*, 321-326.
23. Lerner, M. B.; Resczenski, J. M.; Amin, A.; Johnson, R. R.; Goldsmith, J. I.; Johnson, A. T. C., Toward Quantifying the Electrostatic Transduction Mechanism in Carbon Nanotube Molecular Sensors. *J. Am. Chem. Soc.* **2012,** *134*, 14318-14321.
24. Weiss, J. N., The Hill Equation Revisited: Uses and Misuses. *The FASEB journal : official publication of the Federation of American Societies for Experimental Biology* **1997,** *11*, 835-841.
25. Emmerson, P. J.; Liu, M. R.; Woods, J. H.; Medzihradsky, F., Binding Affinity and Selectivity of Opioids at Mu, Delta and Kappa Receptors in Monkey Brain Membranes. *J. Pharmacol. Exp. Ther.* **1994,** *271*, 1630-1637.
26. Lerner, M. B.; D'Souza, J.; Pazina, T.; Dailey, J.; Goldsmith, B. R.; Robinson, M. K.; Johnson, A. T. C., Hybrids of a Genetically Engineered Antibody and a Carbon Nanotube Transistor for Detection of Prostate Cancer Biomarkers. *ACS Nano* **2012,** *6*, 5143-5149.

27. Cox, R. F.; Collins, M. A., The Effects of Benzodiazepines on Human Opioid Receptor Binding and Function. *Anesthesia and analgesia* **2001,** *93*, 354-358.
28. Carlson, K. E.; Choi, I.; Gee, A.; Katzenellenbogen, B. S.; Katzenellenbogen, J. A., Altered Ligand Binding Properties and Enhanced Stability of a Constitutively Active Estrogen Receptor: Evidence That an Open Pocket Conformation Is Required for Ligand Interaction. *Biochemistry* **1997,** *36*, 14897-14905.
29. Das, A.; Pisana, S.; Chakraborty, B.; Piscanec, S.; Saha, S. K.; Waghmare, U. V.; Novoselov, K. S.; Krishnamurthy, H. R.; Geim, A. K.; Ferrari, A. C*., et al.*, Monitoring Dopants by Raman Scattering in an Electrochemically Top-Gated Graphene Transistor. *Nat. Nanotechnol.* **2008,** *3*, 210-215.
30. Dresselhaus, M. S.; Dresselhaus, G.; Pimenta, M. A.; Malard, L. M., Raman Spectroscopy in Graphene. *Phys. Rep.* **2009,** *473*, 51-87.
31. Kybert, N. J.; Han, G. H.; Lerner, M. B.; Esfandiar, A.; Johnson, A. T. C., Scalable Arrays of DNA-Decorated Graphene Chemical Vapor Sensors. *Nano Res* **2013,** *7*, 95-103.
32. Saby, C.; Ortiz, B.; Champagne, G. Y.; Bélanger, D., Electrochemical Modification of Glassy Carbon Electrode Using Aromatic Diazonium Salts. 1. Blocking Effect of 4-Nitrophenyl and 4-Carboxyphenyl Groups. *Langmuir* **1997,** *13*, 6805-6813.